\documentclass[%
 aip,
 amsmath,amssymb,
preprint,%
]{revtex4-1}

\usepackage{xcolor}
\definecolor{forestgreen(web)}{rgb}{0.13, 0.55, 0.13}

\usepackage{xcolor}

\usepackage{graphicx}
\usepackage{dcolumn}
\usepackage{bm}
\usepackage[utf8]{inputenc}
\usepackage[T1]{fontenc}
\usepackage{mathptmx}
\usepackage{etoolbox}
\usepackage{natbib}
\makeatletter
\def\@email#1#2{%
 \endgroup
 \patchcmd{\titleblock@produce}
  {\frontmatter@RRAPformat}
  {\frontmatter@RRAPformat{\produce@RRAP{*#1\href{mailto:#2}{#2}}}\frontmatter@RRAPformat}
  {}{}
}%
\makeatother
\begin{document}

\preprint{AIP/123-QED}

\title[]{Exploring the generation and annihilation of three dimensional nulls through MHD simulations in initially chaotic field devoid of nulls.}
\author{Yogesh Kumar Maurya}
\affiliation{Udaipur Solar Observatory, Physical Research Laboratory, Udaipur, Rajasthan 313001, India}
\affiliation{Department of Physics, Indian Institute of Technology, Gandhinagar, Gujarat 382055, India}
\email{David.Pontin@newcastle.edu.au, sainisanjay35@gmail.com}
\email{yogeshjn1995@gmail.com, ramit73@gmail.com} 


\author{Ramit Bhattacharyya}
\affiliation{Udaipur Solar Observatory, Physical Research Laboratory, Udaipur, Rajasthan 313001, India}

\author{David I. Pontin}
\affiliation{School of Information and Physical Sciences, University of Newcastle, Callaghan, NSW 2308, Australia}
\author{Sanjay Kumar}
\affiliation{Department of Physics, Patna University, Patna 80005, India}

\date{\today}

\begin{abstract}
Three-dimensional  (3D) magnetic nulls are abundant in the solar atmosphere, as been firmly established through contemporary observations. They are established to be important magnetic structures in, for example, jets and circular ribbon flares. While simulations and extrapolations support this, the mechanisms behind 3D null generation remain an open question. Recent magnetohydrodynamics (MHD) simulations propose that magnetic reconnection is responsible for both generating and annihilating 3D nulls, a novel concept. However, these simulations began with initial magnetic fields already supporting pre-existing nulls, raising the question of whether magnetic reconnection can create nulls in fields initially devoid of them. Previously, this question was briefly explored in a simulation with an initial chaotic magnetic field. However, the study failed to precisely identify locations, topological degrees, and natures (spiral or radial) of nulls, and it approximated magnetic reconnection without fully tracking field line in time. In this paper these findings are revisited in light of recent advancements and tools used to locate and trace nulls, along with the tracing of field lines, through which the concept of generation/annihilation of 3D nulls from chaotic fields is established in a precise manner.
\end{abstract}

\maketitle

Contemporary observations, particularly in the context of circular ribbon flares, unequivocally indicates the existence nulls in the solar atmosphere \cite{2012ApJ...760..101W, 2018ApJ...859..122L, 2020ApJ...899...34L}. Numerous extrapolations and simulations further back this up \cite{2009ApJ...704..485T, 2009ApJ...704..341S, 2021PhPl...28b4502N,mondal2023ApJ...953...84M}. Nevertheless, generation of 3D nulls is still an unresolved problem and merits further attention. Toward this goal, recent magnetohydrodynamics (MHD) simulations by \citet{2023PhPl...30b2901M, maurya2024generation} (hereafter called YRD1 and YRD2) demonstrated magnetic reconnection to be responsible for 3D null generation and their eventual annihilation---indeed a novel suggestion. While YRD1 initiates the simulation with a pre-existing potential null, YRD2 advances the proposal by executing a data-based simulation where the initial magnetic field is extrapolated using vector magnetogram data of a solar active region (AR11977). Although in both the studies null generation was ubiquitous, the initial magnetic field supported pre-existing nulls. A natural question is then whether magnetic reconnection can generate 3D nulls from an initial magnetic field having no such nulls initially. The plausibility of such a scenario has been briefly explored in the simulation by \citet{2020ApJ...892...44N} where the initial magnetic field was chaotic and devoid of any 3D null. Although that study demonstrated the generation of magnetic nulls, it failed to precisely identify their location, topological degree and nature (spiral or radial) using presently available standard tools like the upgraded null detection technique \citet{maurya2024generation}. Moreover, a claim of magnetic reconnection demonstrated by change in field line connectivity requires strict maintenance of the involved magnetic field lines, which was approximated in \citet{2020ApJ...892...44N} by keeping the initial point of field line integration constant at every instant whereas a more precise requirement is to follow the reconnecting field lines as they advect with plasma flow (in the ideal MHD region). For completeness, it is then indispensable to revisit the findings of \citet{2020ApJ...892...44N} in the light of recent understanding and tools developed in YRD1 and YRD2 and put the idea of generation of 3D nulls from chaotic field on a firmer footing.  

Toward this objective, the following presents a brief discussion of the initial magnetic field. The field is constructed by superposing two Arnold-Beltrami-Childress (ABC) fields \cite{ram2014}, each satisfying the linear force-free equation 
\begin{equation}
\label{lfff}
\nabla\times{\bf{B}^\prime}=\lambda{\bf{B}^\prime},
\end{equation}
\noindent having solution 
\begin{eqnarray}
\label{comp}
& & B_x^\prime= A\sin\lambda z + C\cos\lambda y, \\
& & B_y^\prime= B\sin\lambda x + A\cos\lambda z, \\
& & B_z^\prime= C\sin\lambda y + B\cos\lambda x.  
\end{eqnarray}
and, being represented as 
\begin{eqnarray} 
{\bf{B}}={\bf{B}}_1^\prime+d_0{\bf{B}}_2^\prime. 
\end{eqnarray}
The constant $d_0$ relates the amplitudes of the two superposed fields. In Cartesian coordinates the components of $\bf{B}$ are
\begin{eqnarray}
\label{super}
{B_{x}}= A\left(\sin\lambda_1 z + d_0\sin\lambda_2 z \right) + C\left(\cos\lambda_1 y + d_0\cos\lambda_2 y\right) ,\\
{B_{y}}= B\left(\sin\lambda_1 x + d_0\sin\lambda_2 x \right) + A\left(\cos\lambda_1 z + d_0\cos\lambda_2 z\right),\\
{B_{z}}= C\left(\sin\lambda_1 y + d_0\sin\lambda_2 y \right) + B\left(\cos\lambda_1 x + d_0\cos\lambda_2 x\right). 
\end{eqnarray}
Equation (\ref{lfff}) is an eigenvalue equation of the operator $(\nabla\times)$, eigenfunctions of which forms a complete orthonormal basis when eigenvalues $\lambda$ are real \citep{yoshida}. 
Further simplification of (\ref{super}) can be made by selecting $\lambda_1=-\lambda_2=\lambda$, rendering 
\begin{eqnarray}
\label{comph1}
& & {B_{x}}= 0.5A\sin z + 1.5C\cos y ,\\
\label{comph2}
& & {B_{y}}= 0.5B\sin x + 1.5A \cos z ,\\
\label{comph3}
& & {B_{z}}= 0.5C\sin y + 1.5B\cos x,
\end{eqnarray}
for the selection $d_0=0.5$ and $\lambda=1$. The resulting Lorentz force
\begin{eqnarray}
\label{lor}
& &({\bf{J}}\times{\bf{B}})_x=B^2 \sin2 x-2AC\sin y \cos z,\\
& &({\bf{J}}\times{\bf{B}})_y=C^2 \sin2 y-2AB\cos x \sin z,\\
& &({\bf{J}}\times{\bf{B}})_z=A^2 \sin2 z-2BC\sin x \cos y,
\end{eqnarray}
can be utilized to drive the plasma from an initial static state to develop dynamics. Importantly, the ${\bf{B}}$ is chaotic and a detailed discussion can be found in \citet{2017PhPl...24h2902K}
and \citet{2020ApJ...892...44N}. Also, important is the relative magnitudes of the constants A, B and C. For instance, if A=B=1, an increasing C makes the volume occupied by chaotic field larger---a conclusion derived in \citet{2017PhPl...24h2902K}, which can be used as a
measure of chaoticity. For the simulations executed here, notable is the range $0\le C \le 0.3142$, for $A=B=1$, for which $\bf{B}$ is entirely devoid of any magnetic nulls. 
Consequently, using the $\bf{B}$ as an initial condition provides the unique opportunity to explore null generation from a state having no preexisting nulls---the objective of this communication; along with understanding null dynamics in an environment of chaotic magnetic field, left as a future exercise.

The simulations are carried out using the magnetohydrodynamic numerical model EULAG-MHD  \citep{SMOLARKIEWICZ2013608} idealizing the plasma to be thermodynamically inactive, incompressible, and  having perfect electrical conductivity. The governing MHD equations are
\begin{eqnarray}
\label{stokes}
& &\rho_0\left(\frac{\partial{\bf{v}}}{\partial t} 
+ \left({\bf{v}}\cdot\nabla \right) {\bf{ v}} \right) =-\nabla p
+\frac{1}{4\pi}\left(\nabla\times{\bf{B}}\right) \times{\bf{B}}+\nu_0\nabla^2{\bf{v}} ~~~,\\  
\label{incompress1}
& & \nabla\cdot{\bf{v}}=0 ~~~, \\
\label{induction}
& & \frac{\partial{\bf{B}}}{\partial t}=\nabla\times({\bf{v}}\times{\bf{B}}) ~~~, \\
\label{solenoid}
& &\nabla\cdot{\bf{B}}=0 ~~~, 
\end{eqnarray}

\noindent in standard notations and cgs system of units. The constants $\rho_0$ and $\nu_0$ are uniform density and kinematic viscosity, respectively and $\rho_{0}$ represents the constant mass density. Although not strictly applicable in the solar corona, the incompressibility is invoked in other works also \citep{1991ApJ...383..420D, 2005AA...444..961A}. 
With details in \citet{SMOLARKIEWICZ2013608} (and references therein), 
salient features of the EULAG-MHD applicable for this work are summarized here. Crucial to the model is the spatiotemporally second-order accurate, non-oscillatory, forward-in-time, multidimensional, positive-definite advection transport algorithm MPDATA \citep{https://doi.org/10.1002/fld.1071}. The governing prognostic Equations (\ref{stokes}) and (\ref{induction}) are both solved in the Newtonian form with total derivatives of dependent variables and the associated forcings forming the left- and right-hand side, respectively; see Section 2.1 in \citet{SMOLARKIEWICZ2013608} for a discussion. This guarantees identity of null preservation as the associated forcing of the induction equation vanishes at the nulls to the accuracy of the field solenoidality (\ref{solenoid}), which is high \citep{SMOLARKIEWICZ2013608}. 
Another important aspect is the proven dissipative nature of the MPDATA \citep{2003PhFl...15.3890D,2011ApJ...735...46R, 2016AdSpR..58.1538S}. This dissipation is intermittent and adaptive to the generation of under-resolved scales in field variables for a fixed grid resolution. Using this dissipation property, the MPDATA removes under-resolved scales by producing locally effective residual dissipation of the second order in grid increments, enough to sustain the monotonic nature of the solution in advective transport. The consequent magnetic reconnection is then in the spirit of ILESs that mimics the action of explicit subgrid-scale turbulence models, whenever the concerned advective field is under-resolved, as described in \citet{2006JTurb...7...15M}. Such ILESs performed with the model have successfully simulated regular solar cycles by \citet{2010ApJ...715L.133G} and \citet{2011ApJ...735...46R}, with the rotational torsional oscillations subsequently characterized and analyzed in \citet{2013SoPh..282..335B}. The simulations carried out here also utilize the ILES property to initiate magnetic reconnections already shown by \citet{2016ApJ...830...80K}.

The simulations have been carried out for the aforementioned field with $C\in \{0.15, 0.3\}$ to explore null generations with an increase in chaoticity. The kinematic viscosity is set as $\nu = 0.010$, while the spatial and temporal grid increments are $\Delta x = \Delta y = \Delta z = 0.09973$ along the x, y, z axes, respectively and $\Delta t = 0.016$, in CGS units. Triply periodic boundary conditions are applied, and the grid resolution is set to $64 \times 64 \times 64$ mapping a physical dimension of $(2\pi)^3$ to facilitate magnetic reconnection while optimizing the computation costs. Each simulation spans a time of $32$ s. 

\begin{figure}[h]
  \includegraphics[width=\linewidth]{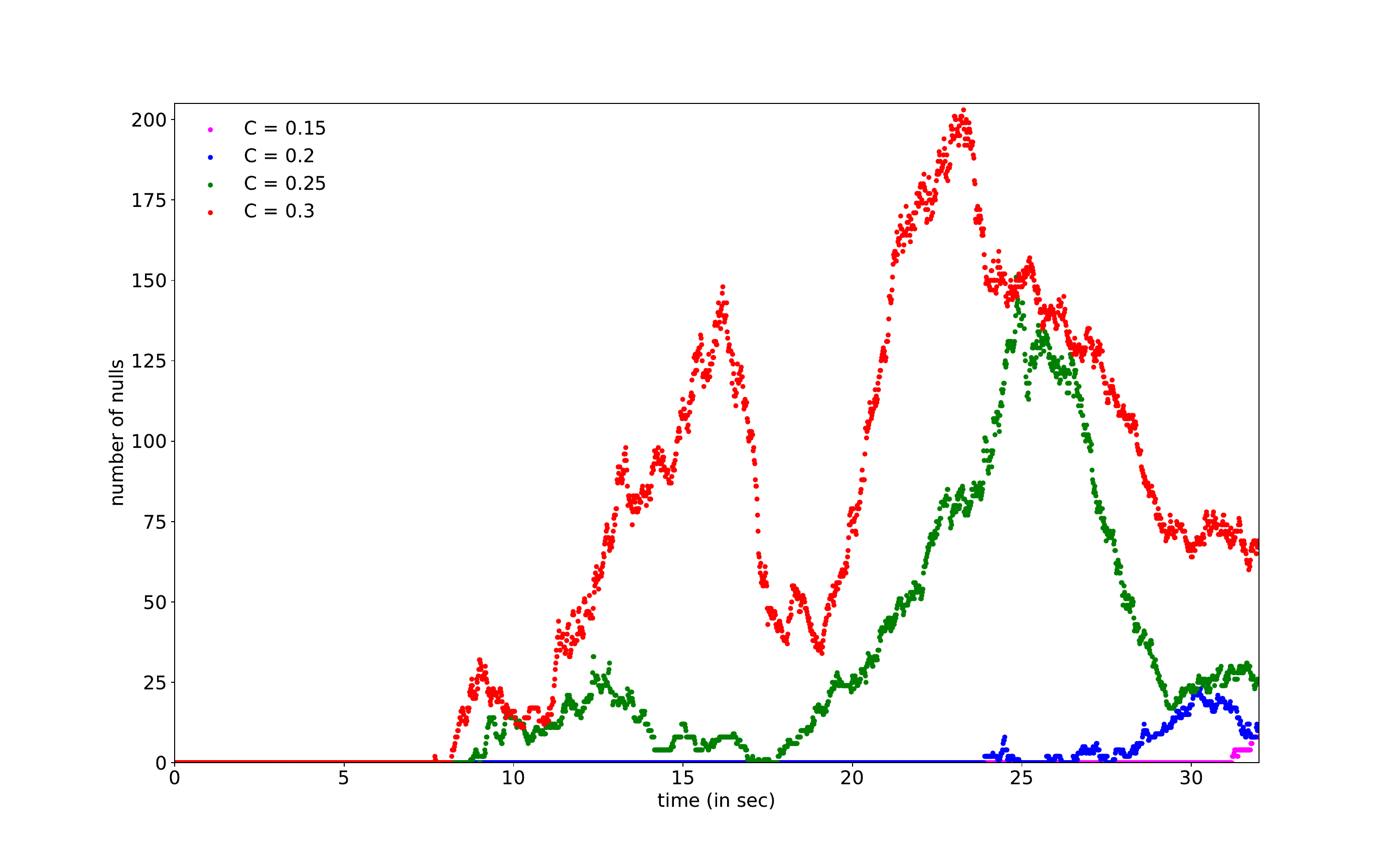}
  \caption{The plot shows an increase in the number of nulls at a given instant and its maximal value over the temporal range with an increase in chaoticity. The vertical axis represents the number of nulls and horizontal axis represents time. The plots in different colors (pink, blue, green, and red) represent the variation in number of nulls for a particular value of the chaoticity (0.15, 0.20,0.25,0.30, respectively). Generation of nulls occur earlier in time as the chaoticity $C$ increases, i.e., $t=(31, 23, 9, 8$)s for $C = (0.15, 0.20, 0.25, 0.30)$.}
  \label{null_var_with_C}
\end{figure}

\begin{figure}
  \includegraphics[width=0.9\linewidth]{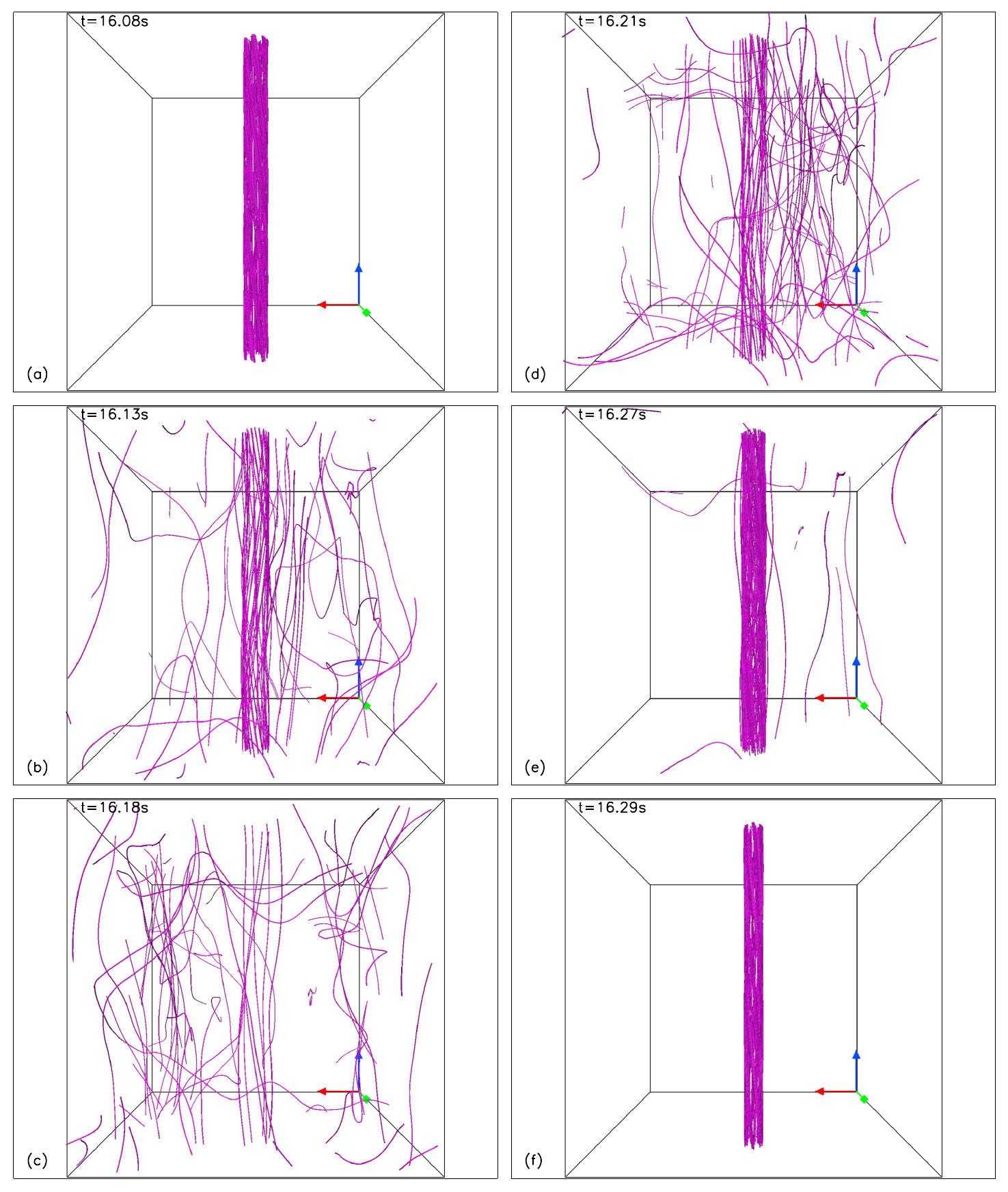}
  \caption{Panels depict a sudden increase in chaoticity near the second peaks in number of nulls, followed by its decrease. In the figure a local flux surface traversed by a single field line for $C=0.3$ in $t\in\{16.08, 16.29\}s$, spanning the second prominent peak at $t=16.18$. Clearly, the surface loses its coherent structure as the line becomes more volume filling and hence chaotic. At $t=16.18s$ (Panel (c)), which marks the second peak, the local flux surface is almost destroyed but reorganized itself at later times (panels (d) to (f)). (An animation is provided as a supplementary material).}
  \label{flux_surface}
\end{figure}

Figure \ref{null_var_with_C} plots the number of nulls with time for different values of $C$, depicting an increase in the number of nulls at a given instant and its maximal value over the temporal range with an increase in chaoticity.  Additionally, nulls appear earlier for larger values of $C$, precisely at $t=(31, 23, 9, 8$)s for $C = (0.15, 0.2, 0.25, 0.3)$. Interestingly, the null generation for all $C$ values is in bursts, most pronounced for $C=0.3$ which shows three identifiable peaks at $t=\{9.26, 16.18, 23.28\}$ seconds. A possible reason can be a sudden increase in chaoticity near the peaks, followed by its decrease. Figure \ref{flux_surface} (Multimedia available online) verifies this ansatz by following a local flux surface traversed by a single field line for $C=0.3$ in $t\in\{16.08, 16.29\}s$, spanning the second prominent peak at $t=16.18s$. Clearly, the surface loses its coherent structure as the line becomes more volume filling and hence chaotic. At $t=16.18s$ (Panel (c)), which marks the second peak, the local flux surface is almost destroyed but reorganized itself at later times (panels (d) to (f)). It has proposed that the presence of chaotic field lines may promote the occurrence of magnetic reconnection in fields without nulls \cite{eyink11,boozer19}, and in this case we see that the increase in chaoticity is contemporal with the generation of nulls and with reconnection (see below). The causal link remains to be fully explored in future investigations. Subsequent retrieval of the flux surface arrests this increase in reconnection----leading to a peak in the number of nulls. Auxiliary analyses (not shown) indicates similar rationale behind all the other peaks also.

\begin{figure}[htbp]
  \includegraphics[width=0.5\linewidth]{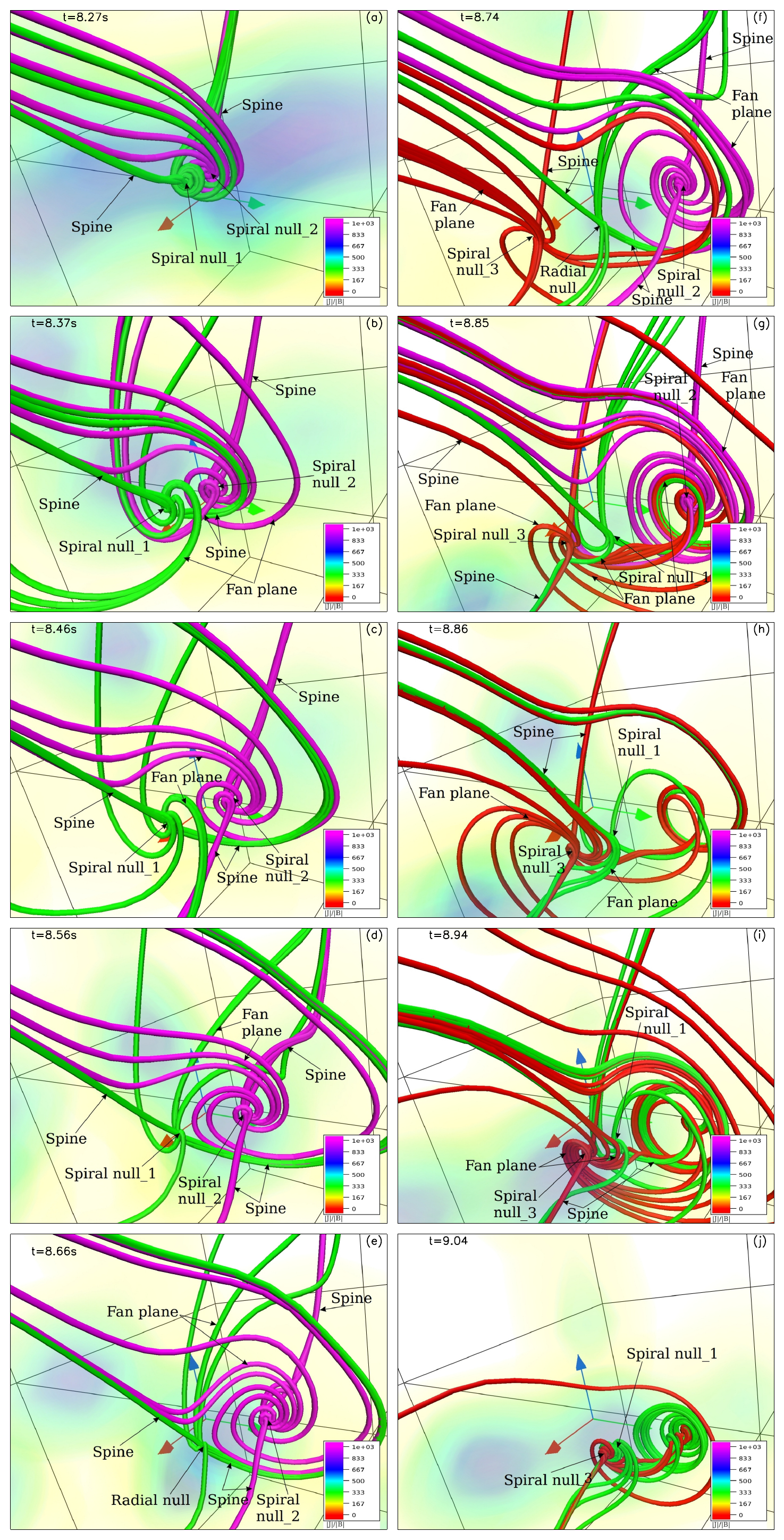}
  \caption{Figure depicts the evolution of nulls with time after their generation. Nulls are traced in time and field lines in green, pink, and red are drawn near the spiral null$\_1$ ($SN1$), spiral null$\_2$ ($SN2$), and spiral null$\_3$ ($SN3$), respectively. The colorbar in panels depicts the magnitude of the $|\bf{J}|/|\bf{B}|$, where $\bf{J}$ and $\bf{B}$ represent current density and magnetic field. Panel (a) depicts the field lines structure near the nulls at $t=8.27$s, the time of their generation, two spiral nulls ($SN1, SN2$) are created. With the evolution $SN1$ and $SN2$ recede away from each other (Panels (b)-(e)) and $SN1$ changes its nature from spiral to radial (panel (e)). Subsequently this radial null reverts back to a spiral null (panels (f)-(h)), which later annihilates with a different spiral null ($SN3$) formed in a distinct null pair generation process (panel (j)).(An animation is provided as a supplementary material).}
  \label{null_traced}
\end{figure}

With chaoticity being directly related to onset of current sheets \citep{2017PhPl...24h2902K} and hence, magnetic reconnections; the results indicate toward the possibility of magnetic reconnection being the underlying reason for the null generations. To further support this hypothesis, a detailed study of field line dynamics leading to the formation and annihilation of nulls is carried out. For this purpose, the dynamics corresponding to $C=0.3$ is selected as the nulls are generated earlier in time and mostly away from the boundaries of the computational domain, leading to their better tractability over time. The focus is set on the nulls generated in a pair with coordinates $(x,y,z) \in \{(0.166, 0.034, 0.101)\pi,(0.169, 0.034, 0.101)\pi\}$, at $t=8.27$s---panel (a) of Fig. \ref{null_traced}) (Multimedia available online) as it involves spiral-spiral pair generation and annihilation, hitherto unexplored in YRD1 and YRD2. Additionally, the pair is created almost at the beginning of the pair generation, being third in the chronology. 

With the experience gained from YRD1 and YRD2, the field lines are advected with the plasma flow and traced in time to reveal the magnetic field line dynamics. The two sets of selected field lines (one in green and the other in pink) are plotted in the ideal regions (i.e., away from reconnection sites) at coordinates $x,y,z \in \{(0.172,0.032, 0.104)\pi,(0.1634, 0.035, 0.106)\pi\}$ and traced in time.

\begin{figure}[htbp]
  \includegraphics[width=0.68\textwidth]{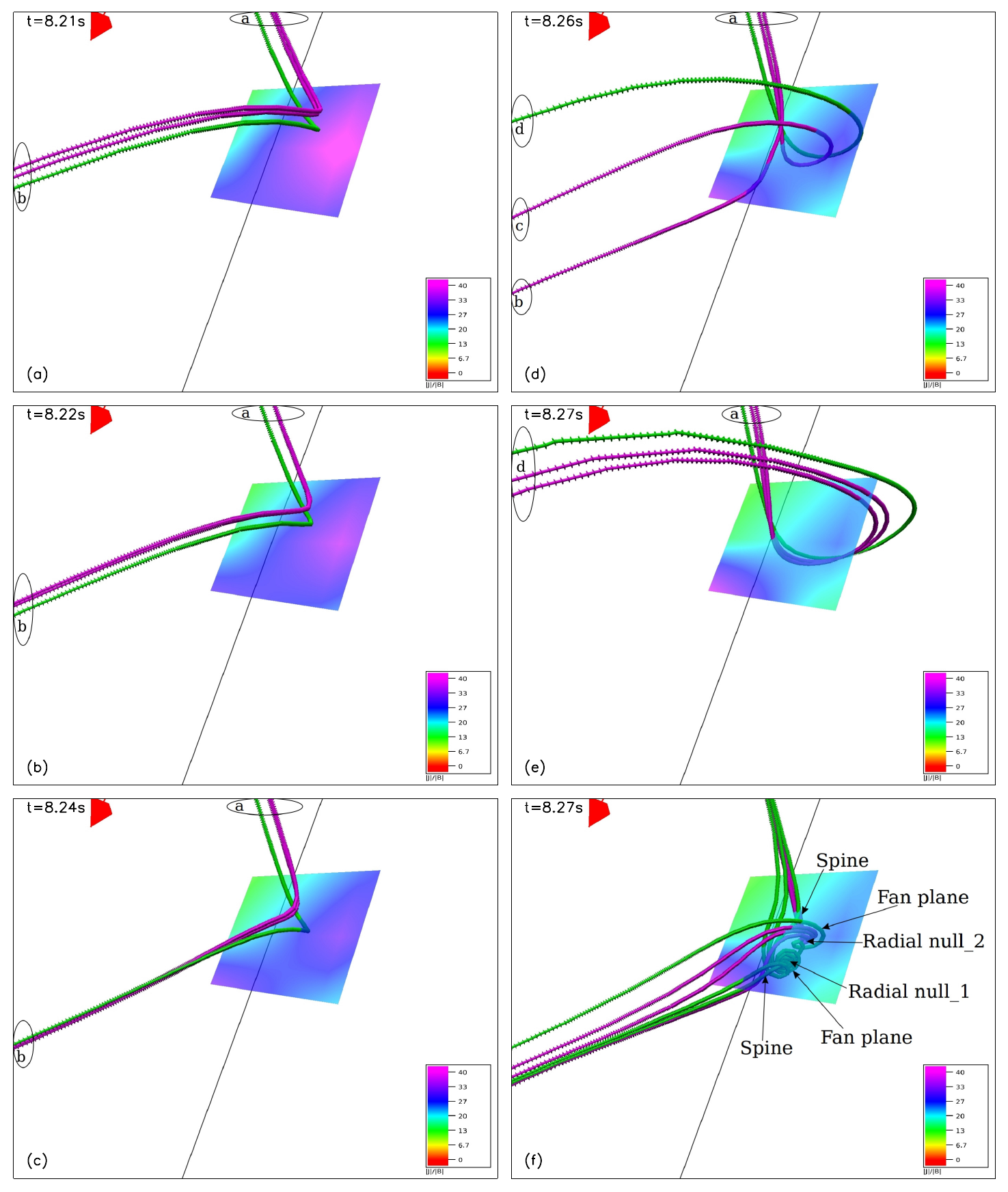}
  \caption{Figure illustrate the magnetic reconnection leading to the spontaneous generation of two spiral nulls. The two pink and one green selected magnetic field lines are drawn in the ideal regions (i.e., away from reconnection sites), advected with the plasma flow, and traced in time. The colorbar represents the same quantity $|\bf{J}|/|\bf{B}|$ as mentioned in the caption of figure \ref{null_traced}. At $t=8.21$s all three (two pink and one green) field lines are connected from region a to region b (panel (a)). With the evolution (acroos panels (a)-(c)), green field line develops more prominent elbow shape and the connectivity of all field lines remain same (i.e. connected from region a to region b) and $|\bf{J}|/|\bf{B}|$ varies accordingly. Across panels (c) to (d), a green and one of the two pink field lines  change their connectivity from regions a to b to regions a to d and c, respectively, which is a telltale sign of magnetic reconnection. With subsequent evolution, the second pink field line also changes its connectivity from regions a to b to regions a to d, while the other pink field lines change their connectivity from regions a to c to the regions a to d (c.f. panels (d)-(e)). Such changes in connectivity are attributed to the magnetic reconnection. The two spiral nulls got spontaneously got generated through this reconnection at $t=8.27$s. The field lines drawn in the vicinity of the generated spiral nulls to show clear structure of both nulls (panel (f)).}
  \label{reconnection_gen_nulls}
\end{figure}

\begin{figure}[htbp]
  \includegraphics[width=0.7\linewidth]{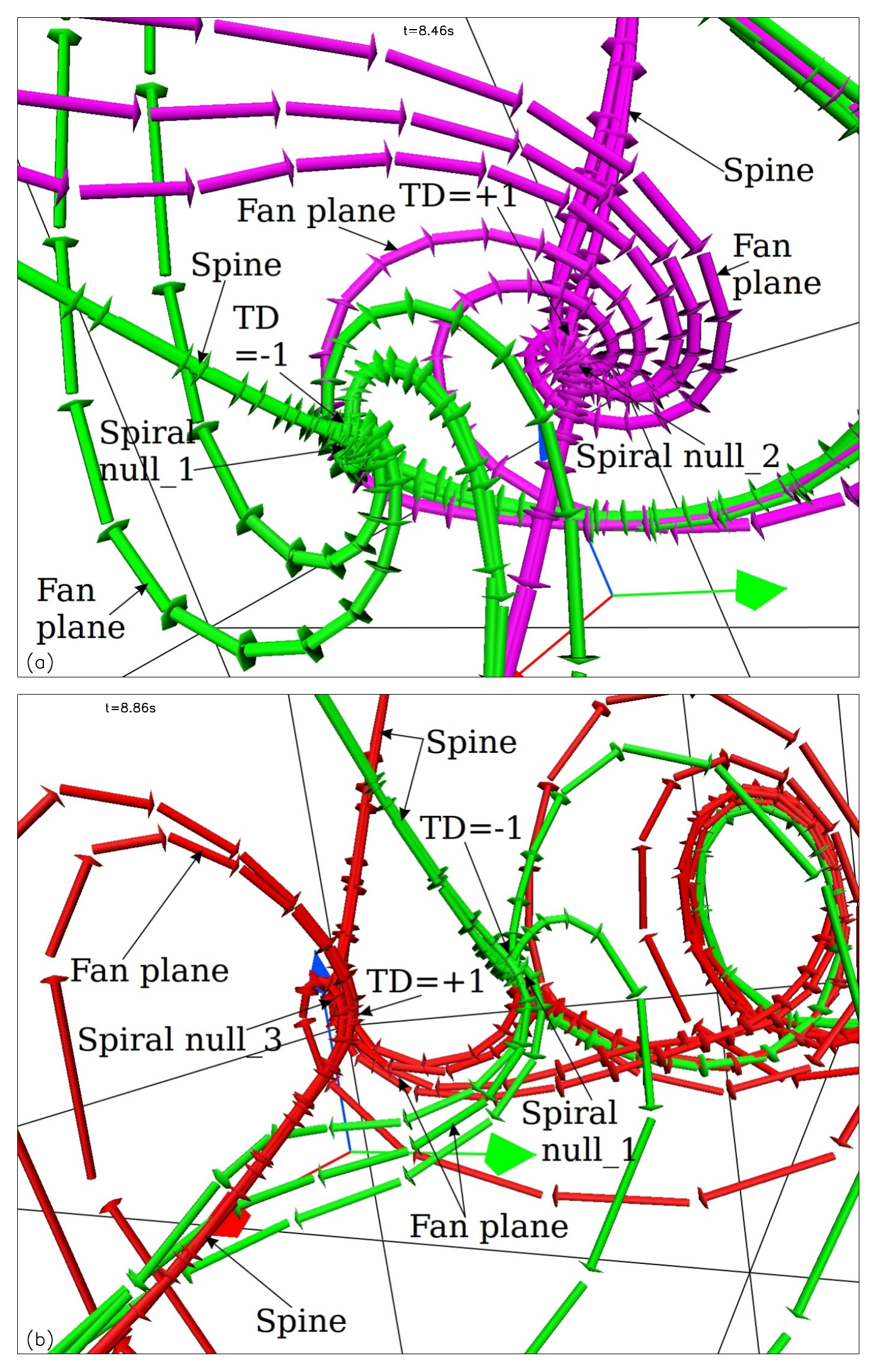}
  \caption{Figure depicts the topological details of the two spontaneously generated spiral nulls at $t=8.46$s (panel (a)) and the null pair at $t=8.86s$ (panel (b)). The fan field lines (in green) of spiral null$\_1$ ($SN1$) are directed away from the null, resulting in a topological degree of $-1$ (panels (a) and (b)), whereas the fan field lines (in pink) of spiral null$\_2$ ($SN2$) are directed towards the null, making a topological degree $+1$. The fan plane field lines of spiral null$\_3$ ($SN3$) (in red) are directed towards the null point making topological degree $+1$ (panel (b)). With time these two nulls get annihilated.}
  \label{two_spiral_null_td}
\end{figure}

At $t=8.26$s, the green and pink field lines are connected from regions a to b (panel (a) of Fig. \ref{reconnection_gen_nulls}) (Multimedia available online). With the evolution, at $t=8.27$s field lines change their connectivity from regions a to b to regions a to c and d, specifically, one green line changes its connectivity from regions a to b to regions a to c, while the other green and pink field lines change their connectivity from regions a to b to the regions a to d. Such changes in connectivity are attributed to magnetic reconnection. The two nulls are generated simultaneously. The eigenvalues of the Jacobian matrix $\nabla{\bf{B}}$ at each null are calculated, and it is found that the imaginary part of the eigenvalues is non-zero for each of the two nulls, implying that they are both spiral nulls (hereafter referred to as $SN1$ and $SN2$, respectively). These nulls are traced in time and field lines are drawn from the close vicinity of the nulls, as shown in Figure \ref{null_traced}. The nulls move away from each other after their generation (c.f. panels (a)-(e)). The topological details of the nulls are illustrated in Figure \ref{two_spiral_null_td} (a). The fan field lines (in green) of $SN1$ are directed away from the null, resulting in a topological degree of $-1$, whereas the fan field lines (in pink) of $SN2$ are directed towards the null, making a topological degree $+1$. The generation of nulls in pair satisfies the conservation of net topological degree. 

With further evolution, the imaginary part of the eigenvalues becomes zero, implying that the $SN1$ loses its spirality and becomes a radial null and remains radial till $t=8.85$s. Subsequently, at $t=8.86$s, the imaginary part of the eigenvalues again becomes non-zero, causing it to revert back to a spiral null $SN1$ (Fig. \ref{null_traced}). Subsequently, $SN1$ approaches another spiral null (panels (f), (g)) and marked as $SN3$), which is one of the spiral-spiral null pair generated earlier at $t=8.1$ and marked as $SN3$ in Fig.\ref{null_traced}. $SN1$, $SN2$, and $SN3$ are traced in time and the green, pink, and red field lines are drawn near $SN1$, $SN2$, and $SN3$, respectively (panels (f)-(g)). $SN1$ and $SN3$ approach each other, and ultimately annihilate pairwise (panels (h)-(j) of Fig. \ref{null_traced}). Similar to Figure \ref{reconnection_gen_nulls}, the annihilation coincides with a change of global field line connectivity (not shown). The spine and fan plane, along with the topological degree, are depicted in Fig. \ref{two_spiral_null_td} (b). The red and green field lines are plotted near $SN3$ and $SN1$. The fan field lines (depicted in red) are directed towards the null, making topological degree $+1$ and the fan field lines (in green) are directed away from the null, resulting in a topological degree $-1$. The conservation of net topological degree is self-explanatory.

This communication relates spontaneous generation/annihilation of 3D nulls with varying levels of chaoticity in an initially chaotic magnetic field while investigating evolution of the involved  magnetic field lines. The initial magnetic fields have been derived by superposing two ABC fields, each satisfying the linear force-force condition. For the computations, $C=0.15, 0.2, 0.25,$ and $0.3$, corresponding to initial fields with increasing chaoticity. The updated trilinear 3D null detection technique \citep{maurya2024generation} has been employed to locate the nulls, calculate their topological degrees, and nature (spiral or radial) based on eigenvalues. These analytically constructed initially chaotic fields do not contain any null (also checked by using null detection technique). The simulation results demonstrate a direct correlation between chaoticity levels and the number of null generations, with higher chaoticity leading to earlier null creations and increased null count. Further to explore null generation/annihilation in more detail, the chaoticity is set at $C=0.3$ as the generation of nulls started earlier in time. As an example of the null generation process, a spontaneously generated pair of spiral nulls is selected. Interestingly, one of the nulls changes its nature from spiral to radial with evolution. Subsequently this radial null reverts back to a spiral null, which later annihilates with a different spiral null formed in a distinct null pair generation process. It is already known that null generation and annihilation require local, non-ideal MHD effects \cite{doi:10.1063/1.871778}. To elucidate the global impact of the creation and annihilation of nulls, the relevant magnetic field lines are traced in time and advected with the plasma flow in the ideal region. It is observed that the field lines change their connectivity from one domain to a different domain---demonstrating that the spontaneous generation (and annihilation) of  3D null point pairs leads to a change in the global field topology.

\begin{acknowledgments}
The computations were performed on the Param Vikram-1000 High Performance Computing Cluster of the Physical Research Laboratory (PRL). SK would like to acknowledge the support from SERB-SURE (SUR/2022/00569). We also wish to acknowledge the visualization software VAPOR (\url{www.vapor.ucar.edu}), for generating the relevant graphics and F. Chiti \emph{et al.} for developing the trilinear null detection technique used here, link of technique is attached (\url{https://zenodo.org/record/4308622#.YByPRS2w0wc}). The corresponding theory can be found in \citet{2007PhPl...14h2107H}.
\end{acknowledgments}

\textbf{Data Availability Statement}

The data that support the findings of this study are available from the corresponding author upon reasonable request.

\bibliography{aipsamp}


\end{document}